\documentclass{PoS}

\usepackage{amsmath,amsfonts,amssymb}
\usepackage{graphicx}
\graphicspath{{./figs/}}

\title{Mass-corrections to double-Higgs production \& TopoID}

\ShortTitle{Mass-corrections to double-Higgs production \& TopoID}

\author{%
  Jonathan Grigo and \speaker{Jens Hoff}%
  \thanks{This work was supported by the DFG through the SFB/TR9
    ``Computational Particle Physics'' and the Karlsruhe School of
    Elementary Particle and Astroparticle Physics (KSETA).  We would
    like to thank Kirill Melnikov and Matthias Steinhauser for the
    productive collaboration and also for numerous cross-checks
    and useful suggestions concerning {\tt TopoID}.}\\
  Institut f\"ur Theoretische Teilchenphysik, Karlsruhe Institute
  of Technology (KIT)\\
  E-mail: \email{jonathan.grigo@kit.edu}, \email{jens.hoff@kit.edu}}

\abstract{%
  We consider power corrections due to a finite top-quark mass $M_t$ to
  the production of a Higgs boson pair within the Standard Model at
  next-to-leading order (NLO) in QCD.  Previous calculations for this
  process and at this precision were done in the limit of an inifinitely
  heavy top quark.  Our results for the inclusive production cross
  section at NLO include terms up to
  $\mathcal{O}\left( 1/M_t^{12} \right)$.\\
  We present the {\tt Mathematica} package {\tt TopoID} which for
  arbitrary processes aims to perform the necessary steps from Feynman
  diagrams to unrenormalized results expressed in terms of master
  integrals.  We employ it for advancing in this process towards
  next-to-next-to-leading order (NNLO) where further automatization is
  needed.}

\FullConference{%
  Loops and Legs in Quantum Field Theory\\
  27 April 2014 - 02 May 2014\\
  Weimar, Germany}

\begin{document}

\section{Introduction}

Its is still an open question whether the scalar particle discovered by
ATLAS and CMS \cite{Aad:2012tfa,Chatrchyan:2012ufa} at CERN is indeed
the Higgs boson of the Standard Model (SM).  In forthcoming years its
couplings to the various gauge bosons and fermions will be measured with
improved precision to verify their compatibility with the values
dictated within the SM.  But to gain insight into the mechanism of
electroweak symmetry breaking the particles self-interactions need to be
probed, too.  The process granting this possibility is production of a
Higgs boson pair via gluon fusion which has two contributions: One where
both Higgs bosons couple to top quarks, the other one involves the cubic
coupling $\lambda$ of the SM Higgs potential (see
fig.~\ref{fig::opt-theo})
\begin{align}
  V(H) = \frac{1}{2} m_H^2 H^2 + \lambda v H^3 + \frac{1}{4} \lambda H^4,
\end{align}
with the Higgs mass $m_H$, vacuum expectation value $v$, and $\lambda =
m_H^2/2 v^2\approx 0.13$ for the SM.  Note that the influence of the
second contribution is strongly suppressed compared to the first one,
but becomes noticeable through its large destructive interference.  The
process has a relatively small cross section and suffers from large
backgrounds, making the extraction of the Higgs self-interaction at the
LHC a challenge.  However, a number of studies suggest the prospect of
measuring $\lambda$
\cite{Baglio:2012np,Baur:2003gp,Goertz:2013kp,Barger:2013jfa}, some
within an accuracy of about 30\% with at least 3000 fb$^{-1}$
accumulated luminosity \cite{Goertz:2013kp,Barger:2013jfa}.

The leading order (LO) result with exact dependence on the top quark
mass $M_t$ has been known since long \cite{Glover:1987nx,Plehn:1996wb}.
Further terms in the perturbation series have been computed in the
approximation of an infinitely heavy top quark $M_t \to \infty$ at NLO
\cite{Dawson:1998py} and just recently at NNLO \cite{deFlorian:2013jea}.
It is important to remark that doing so, the exact LO result has been
factored off in the NLO and NNLO contributions.

\section{Results}

\begin{figure}
  \centering \Large \hspace*{1.0pc}
  \resizebox{0.6\linewidth}{!}{\input{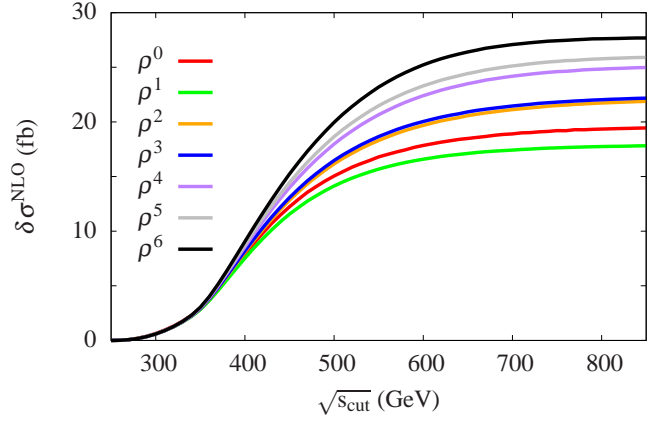}}
  \caption{%
    NLO contribution (without LO) to the hadronic cross section.  The
    color coding indicates higher expansion orders in $\rho =
    m_H^2/M_t^2$.  $\sqrt{s_\text{cut}}$, a cut on the partonic
    $\hat{s}$, can be seen as an approximation for the invariant mass of
    the Higgs pair.}
  \label{fig::xsec-hadr}
\end{figure}

\begin{figure}
  \centering \Large \hspace*{3.0pc}
  \resizebox{0.55\linewidth}{!}{\input{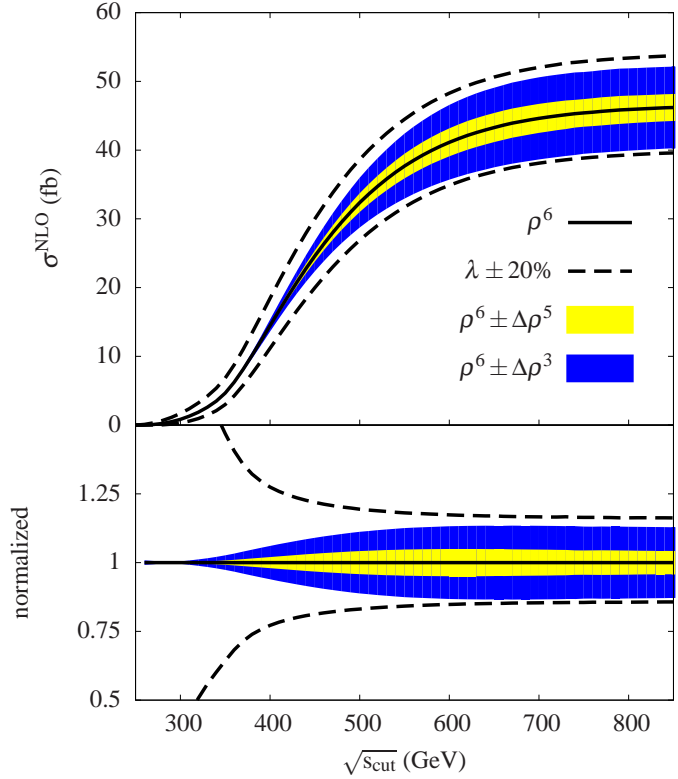}}
  \vspace*{2.5pc}
  \caption{%
    The straight black line shows the hadronic NLO cross section up to
    $\mathcal{O} \left( \rho^6 \right)$, the dashed black lines indicate
    the variation from changing the SM value of $\lambda$ within
    $\pm$20\%.  The yellow and the blue band show the theoretical
    uncertainty by taking the difference to the $\mathcal{O} \left(
      \rho^5 \right)$ and to the $\mathcal{O} \left( \rho^3 \right)$
    expansion, respectively.}
  \label{fig::xsec-lambda}
\end{figure}

It is known that the $1/M_t$ expansion works extremely well for the case
of a single Higgs boson
\cite{Pak:2009dg,Marzani:2008az,Harlander:2009mq} employing the
aforesaid factorization procedure.  For that reason we computed for
double-Higgs production at NLO power corrections due to a finite top
quark mass to the total cross section in the following way:
\begin{align}
  \sigma_\text{expanded}^\text{NLO} \quad \to \quad
  \sigma_\text{exact}^\text{LO}
  \frac{\sigma_\text{expanded}^\text{NLO}}{\sigma_\text{expanded}^\text{LO}},
\end{align}
where numerator and denominator are expanded to the same order in
$1/M_t$.  In \cite{Grigo:2013rya} we presented results expanded up to
$\mathcal{O} \left( 1/M_t^8 \right)$ and in \cite{Grigo:2013xya} to
$\mathcal{O} \left( 1/M_t^{10} \right)$, here they are available to
$\mathcal{O} \left( 1/M_t^{12} \right)$.  The discussion of results has
not changed by including the new terms.  Therefore we only want to show
updated plots for the hadronic cross section, see
fig.~\ref{fig::xsec-hadr} and fig.~\ref{fig::xsec-lambda} and summarize
our findings.

\begin{itemize}
\item The common enhancement by gluon luminosity of low-$\hat{s}$
  contributions, for which we observe good convergence, enlarges the
  validity range of the expansion.
\item Including $1/M_t$ corrections is necessary to detect deviations in
  $\lambda$ of $\mathcal{O} \left( 10\% \right)$.
\item Compared to the prediction in the $M_t \to \infty$ limit we obtain
  for the LHC at 14 TeV
  \begin{align}
    & \sigma^\text{NLO}(p p \to H H) = 19.7^\text{LO} + 19.0^{\text{NLO,
      } M_t \to \infty} \text{ fb} \nonumber\\
    \quad \to \quad & \sigma^\text{NLO}(p p \to H H) = 19.7^\text{LO} +
    (27.3 \pm 5.9)^{\text{NLO, } 1/M_t^{12}} \text{ fb},
  \end{align}
  where no cut on the partonic center-of-mass energy $\hat{s}$ was
  applied and equal factorization and renormalization scale $\mu = 2
  m_H$ was chosen.
\item This can be either seen as an improvement of current precision
  with corrections of about 20\% or at least as reliable error estimate
  for a NLO computation of this process.
\end{itemize}

\section{Techniques}

\begin{figure}
  \begin{align*}
  \begin{aligned}
    \sigma_\text{tot.}\left( gg \to HH \right) \quad & \sim \quad
    \text{Disc.}\left( \mathcal{M} \left( gg \to gg \right) \right)
    \qquad\\[1.0pc]
    \int d \text{PS} \left\vert
      \parbox{60pt}{\includegraphics[scale=0.45]{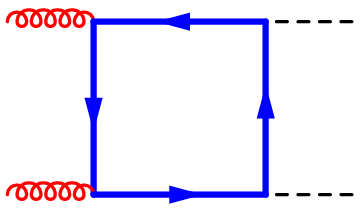}} +
      \parbox{65pt}{\includegraphics[scale=0.45]{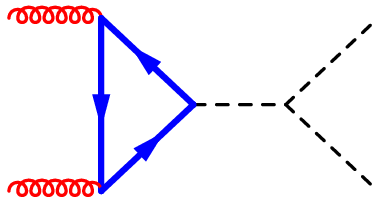}}
    \right\vert^2 \quad & \sim\\[1.0pc]
  \end{aligned}\\
  \begin{aligned}
    \parbox{125pt}{\includegraphics[scale=0.45]{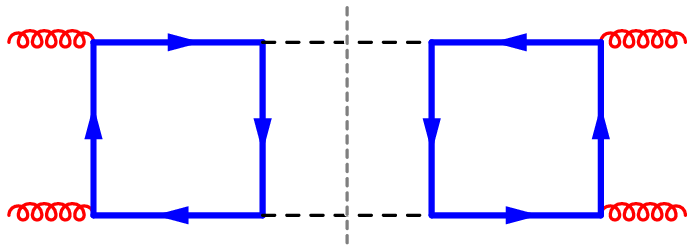}}
    + \parbox{125pt}{\includegraphics[scale=0.45]{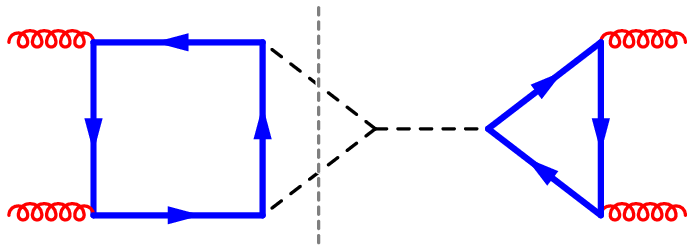}}
    +\parbox{125pt}{\includegraphics[scale=0.45]{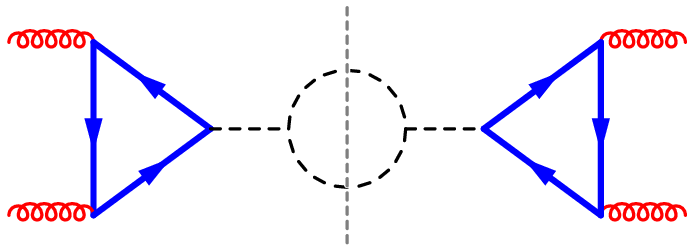}}
  \end{aligned}
  \end{align*}
  \caption{%
    For Higgs boson pair production we only need to consider cuts
    (denoted by short dashed vertical lines) through two Higgs bosons
    (long dashed black lines) and additional partons (beginning at NLO).
    This correspondence is depicted here for the LO order contributions.
    Curly red lines represent gluons, straight blue lines massive top
    quarks.}
  \label{fig::opt-theo}
\end{figure}

Being interested mainly in the total cross section $gg \to HH$, we can
make use of the optical theorem (see, e.g., \cite{Anastasiou:2002yz})
and compute imaginary parts or discontinuities of the amplitude
$\mathcal{M}\left( gg \to gg \right)$ related to a Higgs pair instead of
squaring $\mathcal{M}\left( gg \to HH \right)$ and performing the phase
space integration (see fig.~\ref{fig::opt-theo}).  On the one hand this
method simplifies the calculation, namely: forward scattering
kinematics, common treatment of contributions related to different phase
space integrations and computation of the latter only in the very end at
master integral level.  On the other hand, one has to compute a larger
number of diagrams with more loops.

\begin{figure}
  \centering
  \vspace*{11pt}
  \raisebox{-11pt}{\parbox{120pt}{
      \centering
      \includegraphics[scale=0.7]{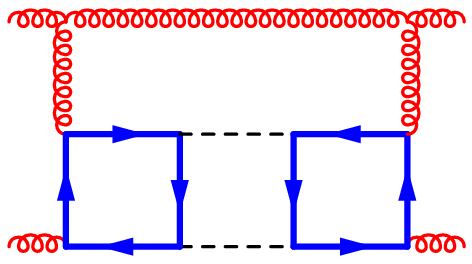}\\
      \vspace*{0.5pc}
      $(M_t^2, m_H^2, s)$
    }}
  $\longrightarrow$
  \raisebox{-11pt}{\parbox{40pt}{
      \centering
      \includegraphics[scale=0.7]{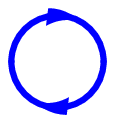}\\
      \vspace*{0.5pc}
      $(M_t^2)$
    }}
  $\times$
  \raisebox{-11pt}{\parbox{100pt}{
      \centering
      \includegraphics[scale=0.7]{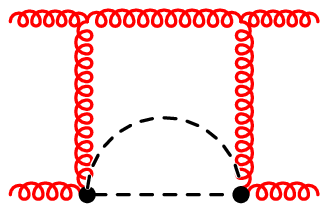}\\
      \vspace*{0.5pc}
      $(m_H^2, s)$
    }}
  $\times$
  \raisebox{-11pt}{\parbox{40pt}{
      \centering
      \includegraphics[scale=0.7]{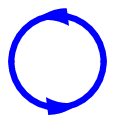}\\
      \vspace*{0.5pc}
      $(M_t^2)$
    }}
  \vspace*{1.0pc}
  \caption{%
    Applying the rules for asymptotic expansion to a single Feynman
    diagram one obtains in general a sum of contributions (there is only
    one in this example).  Each contribution in turn is a product of
    subgraphs (containing the hard scale; $M_t^2$ in our case) and
    co-subgraphs (containing the soft scales; $m_H^2$, $s$).  The
    notation is as in fig.~\protect\ref{fig::opt-theo}.}
  \label{fig::asy-exp}
\end{figure}

The second ingredient making this calculation feasible is the asymptotic
expansion at diagrammatic level (see, e.g., \cite{Smirnov:2002pj}) in
the hierarchy $M_t^2 \gg \hat{s}, m_H^2$ which corresponds to a series
expansion of an analytic result in the parameter $\rho = m_H^2/M_t^2$.
This procedure effectively reduces the number of loops and scales in the
integrals to be evaluated (see fig.~\ref{fig::asy-exp}), thus
diminishing some of the drawbacks connected to use of the optical
theorem.

Within this framework our toolchain for the various steps of the
calculation looks as follows:
\begin{enumerate}
\item generation of Feynman diagrams with {\tt
    QGRAF}~\cite{Nogueira:1991ex},
\item selection of diagrams which have the correct cuts
  \cite{unpublished}, \label{enum::filter}
\item asymptotic expansion with {\tt q2e} and {\tt
    exp}~\cite{Harlander:1997zb,Seidensticker:1999bb},
  \label{enum::asyexp}
\item reduction to scalar integrals in {\tt
    FORM}~\cite{Vermaseren:2000nd,Kuipers:2012rf} and/or {\tt
    TFORM}~\cite{Tentyukov:2007mu},
  \label{enum::scalar}
\item reduction to master integrals by {\tt rows}~\cite{unpublished}
  and/or {\tt FIRE}~\cite{Smirnov:2008iw,Smirnov:2013dia},
  \label{enum::reduce}
\item minimization of the set of master
  integrals~\cite{unpublished}. \label{enum::minimize}
\end{enumerate}
Step~\ref{enum::filter} is necessary since one cannot steer {\tt QGRAF}
in such a way that only diagrams with a specific cut structure are
generated.  Because of that we filter the diagrams provided by {\tt
  QGRAF} for those which exhibit an appropriate cut in the s-channel
corresponding to an interference term from squaring the amplitude for
$gg \to HH$.  (Usually only about 10-30\% of the initial diagrams pass
the filter.)  At NLO step~\ref{enum::scalar} turned out to be the
bottleneck of the calculation for going to higher orders in the
expansion parameter $\rho$.

\section{TopoID}

Up to now the input for steps~\ref{enum::asyexp}-\ref{enum::minimize} in
the above list was usually provided manually.  For going beyond NLO we
use {\tt TopoID} to provide all that information in an automatic
fashion.  More precisely: all the graphs corresponding to a topology as
``mapping patterns'' for step~\ref{enum::asyexp}, {\tt FORM} code
processing aforementioned topologies in step~\ref{enum::scalar} and
definitions of topologies suitable for reduction with the programs
listed for step~\ref{enum::reduce}.

When performing a multi-loop calculation one often works with a set of
topologies and within each topology integrals are reduced to a finite
set of master integrals.  The same master integral may thus be
represented in different ways by single integrals of various topologies.
{\tt TopoID} is capable of providing such an identification, as are
recent versions of {\tt FIRE}~\cite{Smirnov:2013dia}.  Moreover, there
exist also non-trivial \emph{linear} relations involving multiple
``master integrals'' which can be found with the help of this package
(step~\ref{enum::minimize}).

\begin{figure}
  \centering
  \includegraphics[scale=0.8]{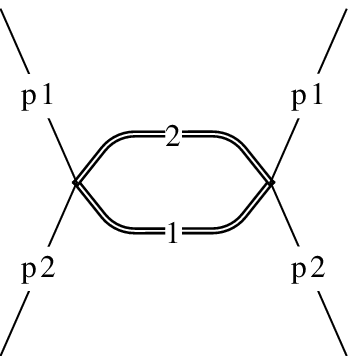}\hspace*{1.5pc}
  \includegraphics[scale=0.8]{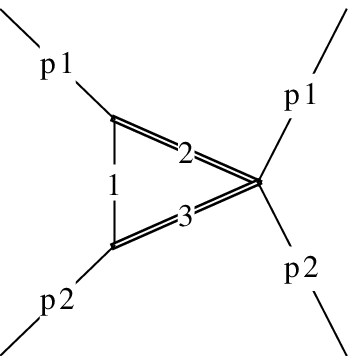}\hspace*{1.5pc}
  \includegraphics[scale=0.8]{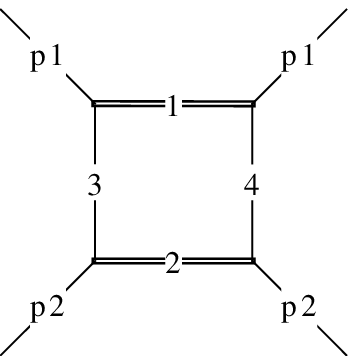}\hspace*{1.5pc}
  \includegraphics[scale=0.8]{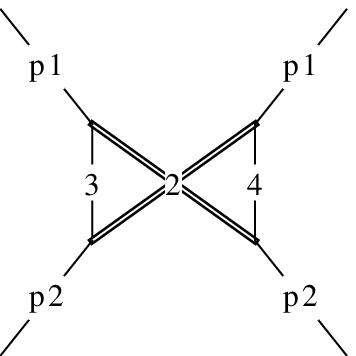}
  \caption{%
    Sample one-loop topologies appearing after asymptotic expansion at
    LO, NLO and NNLO (the last two).  The first graph is an example of a
    linearly independent, but incomplete topology.  The second topology
    is a linearly independent and complete.  The last two topologies,
    one planar, one non-planar, are linearly dependent and complete.
    Plain black lines are massless, the double lines carry the Higgs
    mass.}
  \label{fig::top-type}
\end{figure}

A diagram class or family $T$, usually referred to as topology, is a set
of $N$ scalar propagators $\left\{ d_i \right\}$ with arbitrary powers
$\left\{ a_i \right\}$, usually referred to as indices, composed of
masses $\left\{ m_i \right\}$ and line momenta $\left\{ q_i \right\}$.
The line momenta $\left\{ q_i \right\}$ are linear combinations of $E$
external momenta $\left\{ p_i \right\}$ and $I$ internal momenta
$\left\{ k_i \right\}$ with integers $c_{ij}$, $d_{ij}$,
\begin{align}
  T \left( a_1, \ldots, a_N \right) = \left\{ \prod_{i = 1}^I \int
    dk_i^D \right\} \left\{ \prod_{j = 1}^N \frac{1}{\left[ m_j^2 +
        q_j^2 \right]^{a_j}} \right\},\\
  q_i = \sum_{j = 1}^E c_{ij} p_j + \sum_{j = 1}^I d_{ij} k_j.
\end{align}
For particular kinematics, i.e. given external and internal momenta,
supplemented by possible constraints, e.g. putting particles on-shell,
one can form all occuring scalar products
\begin{align}
  x_{p_i p_j} = p_i \cdot p_j,\quad s_{p_i p_j} = p_i \cdot p_k,\quad
  s_{k_i k_j} = k_i \cdot k_j.
\end{align}
If the denominators of a topology $\left\{ d_i \right\}$ allow for
expressing each of the internal scalar products $s_{ij}$ the topology is
complete, otherwise incomplete.  In the latter case affected scalar
products are called irreducible scalar products and appear only as
numerators (fig.~\ref{fig::top-type} shows some examples for Higgs pair
production).

\begin{figure}
  \centering
  \includegraphics[scale=0.6]{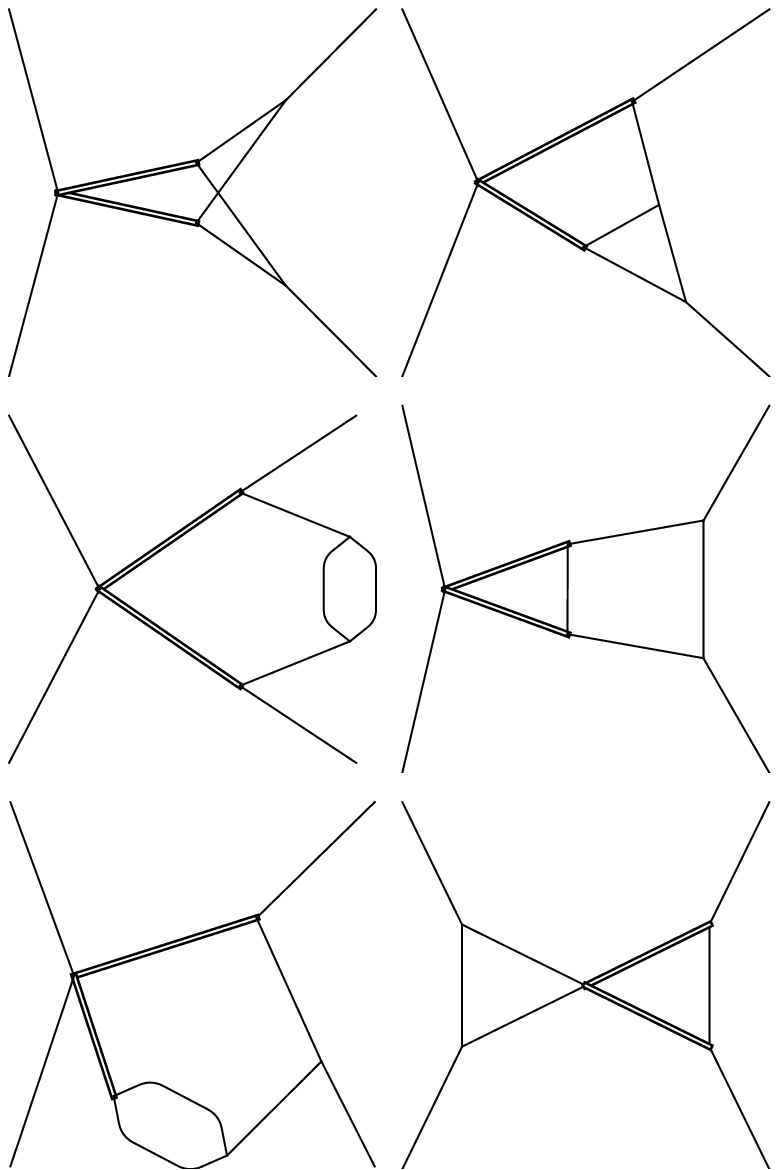}\hspace*{3.5pc}
  \includegraphics[scale=0.6]{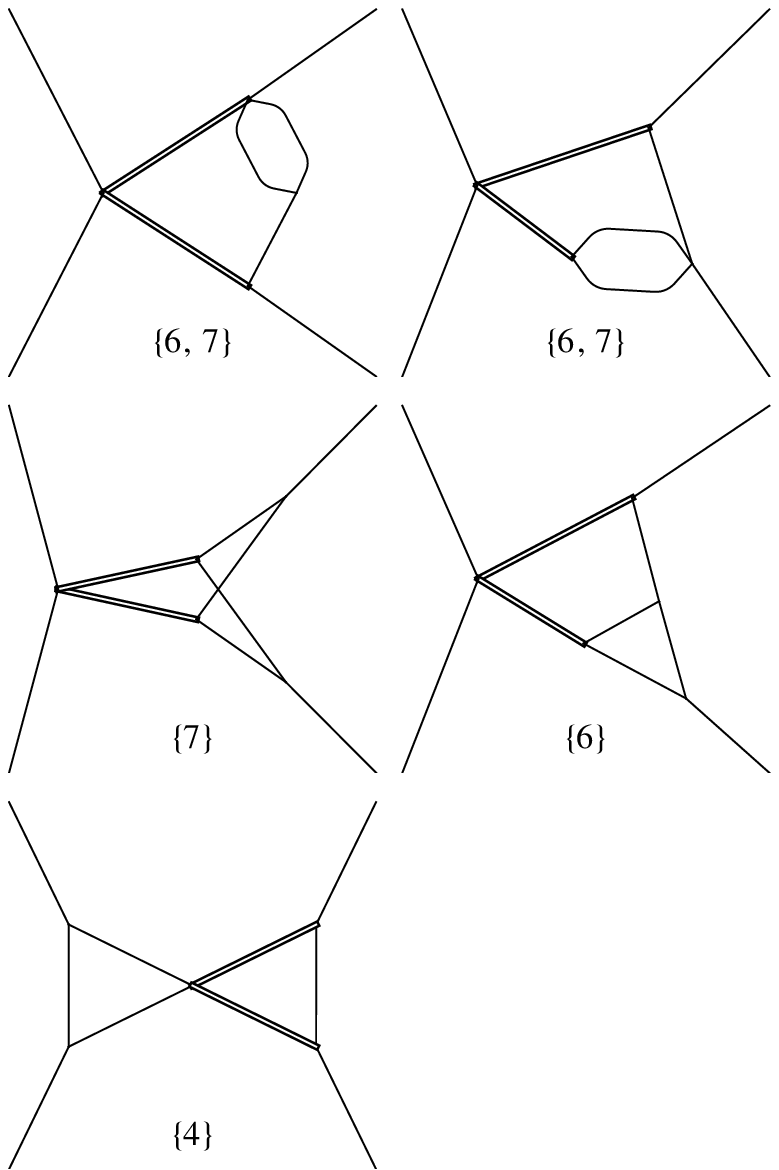}
  \caption{%
    The left hand side shows the set of diagram topologies, the right
    hand side the set of reduction topologies.  Their order is chosen by
    {\tt TopoID} in a fixed way, numbers in braces denote the presence
    of irreducible numerators.  The last topology in both sets is an
    example of a factorizing topology.  Note the modification of
    self-energy insertions from the left to the right side, propagators
    carrying the same momentum are identified.  Furthermore, there is a
    non-trivial mapping from the second and fourth diagram topology to
    the fourth reduction topology which cannot be deduced from the
    graphs alone.}
  \label{fig::tops-NNLOgv22}
\end{figure}

Diagram topologies, i.e. mapping patterns for Feynman diagrams, in
general are incomplete and also linearly dependent, viz. linear
relations among the $\left\{ d_i \right\}$ exist.  In contrast,
reduction topologies need to be linearly independent and complete.  This
is exemplified in fig.~\ref{fig::tops-NNLOgv22} with the two-loop
topologies emerging after asymptotic expansion of the purely virtual
five-loop diagrams at NNLO\footnote{%
  In this case two massive tadpole diagrams containing the top quarks
  (one with one loop and one with two loops) and a two-loop box diagram
  with Higgs mass remain after asymptotic expansion.}.  The mapping
between these two types of topologies can in general become quite
intricate for larger sets but is handled easily by {\tt TopoID}.

The foundation of this automatization is the $\alpha$-representation of
Feynman integrals
\begin{align}
  T\left( a_1, \ldots, a_N \right) = c \left\{ \prod_{i = 1}^N
    \int_0^\infty d \alpha_i \right\} \delta \left( 1 - \Sigma_{i = 1}^N
    \alpha_i \right) \left\{ \prod_{j = 1}^N \alpha_j^{a_j} \right\}
  \mathcal{U}^a \mathcal{W}^b,
\end{align}
where $c$, $a$ and $b$ depend on $I$, $D$ and the $\left\{ a_i \right\}$
only.  The polynomials $\mathcal{U}$ and $\mathcal{W}$ are homogeneous
in the $\left\{ \alpha_i \right\}$ and encode the complete information
on the topology (for further details see, e.g.,
ref.~\cite{Smirnov:2012gma}).  This representation is unique up to
renaming of the $\alpha$-parameters, but this ambiguity can be
eliminated by applying the procedure described in \cite{Pak:2011xt} to
derive a canonical form of the $\alpha$-representation, making it a
suitable identifier.

{\tt TopoID} is a generic, process independent tool and bridges the gap
between Feynman diagrams and unrenormalized results expressed in terms
of \emph{actual} master integrals, i.e. including the non-trivial
relations, in a completely automatic way.  It is written as a package
for {\tt Mathematica} which offers a high-level programming environment
and the demanded algebraic capabilities.  However, for the actual
calculation {\tt FORM} code is generated to process the diagrams in an
effective way.  Let us briefly summarize features the package has to
offer:
\begin{itemize}
\item topology identification and construction of a minimal set of
  topologies,
\item classification of distinct and scaleless subtopologies,
\item access to properties such as completeness, linear dependence,
  etc.,
\item construction of partial fractioning relations,
\item revealing symmetries (completely within all levels of
  subtopologies),
\item graph manipulation, treatment of unitarity cuts, factorizing
  topologies,
\item {\tt FORM} code generation (diagram mapping, topology processing,
  Laporta reduction),
\item master integral identification (arbitrary base changes,
  non-trivial relations).
\end{itemize}
As one cross-check we repeated the NLO calculation within this
automatized setup and found agreement with our previous calculation.

\end{document}